# Some results of CCD-photometry of variable stars at the Astronomical Institute of Karazin Kharkiv National University

V. G. Shevchenko[1,2], D. O. Danylko[2], I. G. Slyusarev[1,2], R. A. Chigladze[3]

[1]Institute of Astronomy of V. N. Karazin Kharkiv National University, Kharkiv, Ukraine
[2]Department of Astronomy and Space Informatics of V. N. Karazin Kharkiv National University, Kharkiv, Ukraine
[3]Samtskhe-Javakheti State University, Kharadze Abastumani Astrophysical Observatory, Georgia

Email: shevchenko@astron.kharkov.ua

**Abstract**

We presented photometric observations for the one UV Ceti type and three W Ursae Majoris- type variable stars. The flare of the UV Ceti type star lasted about two hours, and the star changed magnitude to 3.9 within about two minutes. The values of color indices *V-R*, the rotational periods and the composite lightcurves have been obtained for the EW stars. Using a relation of an absolute magnitude- period obtained by Mateo and Rucinski (2017) and interstellar extinction from the three-dimensional map of Milky Way dust (http://argonaut.skymaps.info) and Green et al. (2019), we have calculated the absolute magnitudes of the EW stars and distances to them. The parallaxes obtained from our data differ from those given in Gaia DR 3, which may be due to insufficient quality calibration of the absolute magnitude-period relation and with the estimations of interstellar extinction.

## 1. Introduction

Many of different types of variable stars were discovered last times due to special ground-based and space-based survey programs, especially Gaia (Ansari et al. 2023, Chen et al. 2020, Gavras et al. 2023, Mowlavi et al. 2023. Qian et al. 2017, Rimoldini, et al. 2023, Samus et al. 2017, etc.). This made it possible to significantly improve statistics on the number of variable stars in the Milky Way, to identify some correlations between the period of brightness changes and absolute magnitude, and to determine the general location of variables of different types in our Galaxy. Despite these achievements, the total number of some types of variables is still not sufficient to clarify their nature of variability. An example would be flaring or pulsating stars, whose variability time is about one-two hours. Checking star fields in different observing programs allows to detect new variable stars or to get a new information about already discovered stars. In our asteroid programs (Chiorny et al. 2023, Shevchenko et al. 2021, 2022), we have checked star fields for star variability and we discovered several variable stars of different types (Shevchenko et al. 2013, 2018, Zheleznyak et al. 2015). In this paper, we present new observational results of some variable stars of different types including one new detecting flaring star.

## 2. Observations and Results

We have carried out the CCD-observations of variable stars in 2013 - 2015 for 15 nights with the 0.7-m reflector at the Chuguyiv Observational Station of the Astronomical Institute of V.N. Karazin Kharkiv National University as a part of student astrophysical practice (Shevchenko et al. 2013, 2018) and the some programs of an investigation of the asteroids (Shevchenko et al. 2021, 2022). Observations on this telescope were stopped on February 24, 2022 because of russian invasion to Ukraine. The observation station was occupied and barbarously destroyed by russian troops. Unfortunately, the telescope is now inactive. The telescope was equipped with a ML4710 CCD camera (1056×1027 pixels, pixel size 13×13 mkm). The camera's field of view was 16.2′×16.2′ with the scale 0.96″/pixel. The quantum efficiency of the ML 4710 in the wavelength range of 0.55 to 0.75 mkm is 93 - 95%, which makes it possible to perform highly efficient and high-quality photometric observations of faint objects. Images were taken using *V* and *R* bands of the Johnson–Cousins photometric system. Original CCD-frames were calibrated for dark current and flat field in the standard manner. The CCD-observations and data reduction methods were explained by Krugly et al. (2002) and by Chiorny et al. (2023). The brightness measurements of stars on CCD-images were done using the aperture photometry package (ASTPHOT) developed by Mottola et al. (1995). The absolute calibrations of the magnitudes were performed using the comparison stars in the SDSS (*ugriz*) photometric system, taken from the APASS DR9 (Henden et al. 2012) and Pan-STARRS DR1 (Chambers et al. 2016) catalogs. To transform them to the Johnson-Cousins (*UBVRI*) photometric system, we used the corresponding equations given in Fukugita et al. (1996), and in Tonry et al. (2012). The accuracy of the resultant absolute photometry is within 0.01-0.03 mag. The number of the star in according to the catalog Gaia DR 3, the equatorial coordinates at the epoch J2000.0, the variable type, the maximum value of apparent magnitude in *R* band, and average color index *V-R* are listen in the Table 1. In section 2.1, we present short characteristics, lightcurve, and some physical parameters for the flare star. The results of observations of three W Ursae Majoris-type binaries (EW type) are given in section 2.2. Some physical characteristics of EW stars are presented in Table 2: number of star in catalogue, rotational period, average amplitude, parallax in milliarcseconds from Gaia ESA Archive (https://gea.esac.esa.int/archive/), absolute magnitude, interstellar extinction and calculated parallax. Some conclusions we resume in section 3.

### 2.1. Flaring star

Flaring stars or UV Cet stars are a subclass of irregular variables, a type of eruptive stars of the dMe spectral class. They are red dwarfs, so the mass of such stars varies from 0.04 to 0.16 solar masses, and the linear dimensions of red dwarfs are three times smaller than the diameter of the Sun (Gershberg et al. 1999). A distinctive feature of these stars is a sharp and non-periodic increase in luminosity, which can reach up to $6 \cdot \times 10^{31} \div 4 \cdot \times 10^{33}$ erg/s (Kowalski 2024). The lightcurve reaches its maximum extremely quickly (from dozens of seconds to several minutes), while the loss of brightness takes longer (from tens minutes to several hours). The amplitude of a flash is also not the same for all stars of this type. In addition to the main increase in brightness, the lightcurve may contain such structural components as small flares before and after the main maximum. The General Catalogue of Variable Stars (Samus et al. 2017) contains just over 1000 stars of this type.

Our observations of the flaring star Gaia DR 3 28446770992650624 were performed only one night on September 3, 2016 in *R* Johnson-Cousins band. The lightcurve is pictured in fig. 1. Beginning moment of flare is JD0 2457634.48130. Magnitude of star during two minutes has changed from 19.2 up to 15.3, after this the magnitude changed down to 18.7 for two hours. In Gaia ESA Archive (https://gea.esac.esa.int/archive/) the parallax of this star is 4.905 ± 0.271 milliarcseconds and the photometric distance was determined to be 198 ± 6 parsec. There have been no previous reports of flare activity about this star.

### 2.2. EW binary stars

*Gaia DR 3 1994611232070701184.* We observed of this system in 2013 for two nights. The composite lightcurve is presented in fig. 2. The beginning moment of observations is JD0 2456491.27959. The rotation period is 0.36972 ± 0.00005 days and maximum amplitude equals 0.42 ± 0.02 mag with different in depth of minima on 0.06 mag, as it can see from fig. 2, that points out on differences in the star luminosities. Color index is equal to *V-R* = 0.47 ± 0.02 mag in maximum and *V-R* = 0.53 ± 0.03 mag in minimum. In Gaia ESA Archive, the photometric distance to this system is provided to be equal to 989 ± 50 parsec. Our value of rotational period is different from those determined by Heinze et al. (2018) (0.3687 day), but their value was obtained from sparse data, that limits the accuracy of determination of the period.

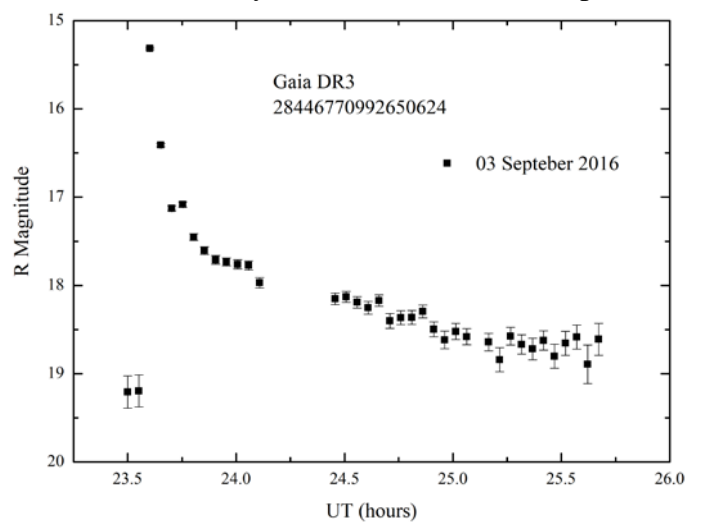

Fig. 1. Lightcurve of flaring star Gaia DR 3 28446770992650624.

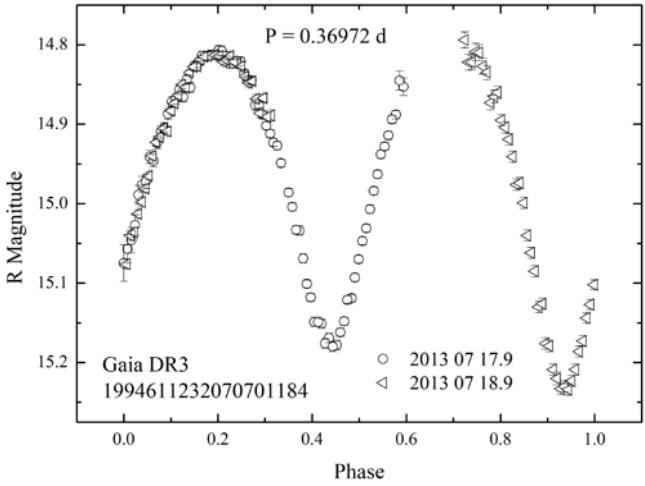

Fig. 2. Composite lightcurve of binary star Gaia DR3 1994611232070701184.

*Gaia DR 3 1997506143105623808.* The observations of this star were carried out during four nights in July-August 2014. Composite lightcurve obtained with the period of 0.408354 ± 0.000015 days is shown in fig. 3. The beginning moment is JD0 2456858.36812. The value of the period is different from those obtained by Chen et al. (2020) (0.4084952 d), but their value was obtained from sparse data, that limits the accuracy of determination of the period. The amplitude of lightcurve equals 0.42 mag. Average color index *V-R* is equal to 0.48 ± 0.03 mag. In Gaia ESA Archive, the photometric distance to this system is provided to be equal to 1560 ± 55 parsec.

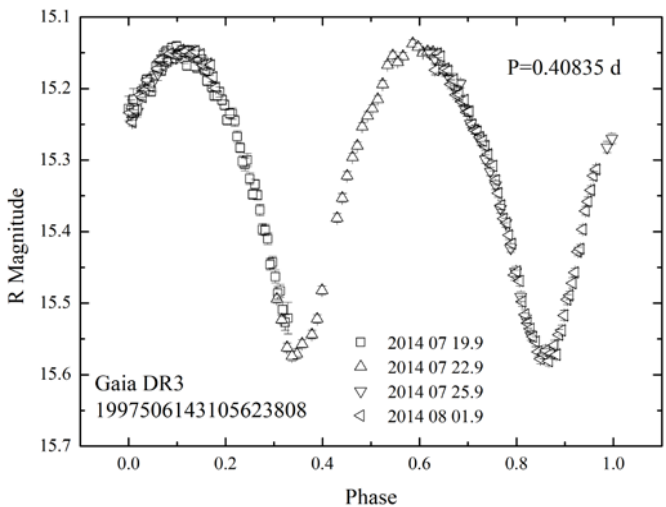

Fig. 3. Composite lightcurve of binary star Gaia DR3 1997506143105623808.

*Gaia DR 3 1997847472748694272.* The observations of this star were carried out for two nights in July 2013 and for six nights in July 2015. A large range of observing time allowed us to determine the rotational period of the system with high precision that is equal to 0.3381355 ± 0.0000015 days, the beginning moment is JD0 2456493.29010. Our value of the period is different from obtained by Chen et al. (2020) (0.338141). Composite lightcurve (fig. 4) has the maximum amplitude of 0.34 ± 0.02 mag and the neighboring minima differ on 0.05 ± 0.01 mag. There are small differences in lightcurves obtained in 2013 and in 2015, which may be related to physical processes in a close binary system. Average color index *V-R* is equal to 0.56 ± 0.03 mag. It should be noted that there is a field star near the variable star, which was not resolved by our aperture photometry. Therefore, the contribution of the field star was included in the estimates of the apparent magnitude of the variable star. Based on this, we did not determine the parallax of the star from our photometric data.

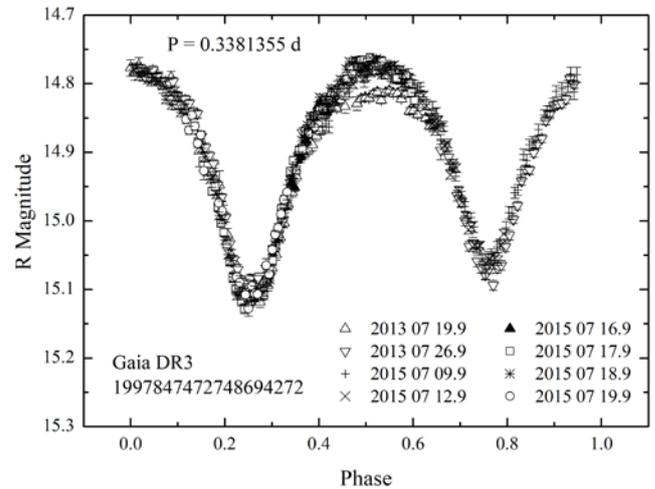

Fig. 4. Composite lightcurve of binary star Gaia DR3 1997847472748694272.

Mateo and Rucinski (2017) have performed a calibration to determine absolute magnitudes $M_V$ for 297 contact EW binary systems with $0.275 < P < 0.575$ days using Tycho-Gaia Astrometric Solution parallax data. They have obtained the next relation between period and absolute magnitude: $M_V = 3.73$ (±0.06) $- 8.67$ (± 0.65) × ($\log(P) + 0.40$). We have used also the three-dimensional map of Milky Way dust (http://argonaut.skymaps.info), and the coefficient to convert $E(g-r)$ into extinction in *V* band from Green et al. (2019) for a determination of the interstellar extinction. Using the absolute magnitude-period relation and the interstellar extinction values, we have calculated the absolute magnitudes and the distances to the stars. Our results are presented in Table 2 (excluding third star). As it can see from Table 2, there are significant differences between Gaia data and our estimations of photometric parallaxes. It can be connect both with the calibration equation and with the values of interstellar extinction for these stars.

## 3. Conclusions

Thus, processing of photometric observations of star fields obtained for 15 nights allowed us to detect a new UV Cet flaring star and to investigate more detail three EW binary stars. The flaring of the UV Ceti type star has continued about two hours and the star has changed its magnitude on 3.9 mag during about two minutes. The values of color indices *V-R* were obtained, the rotational periods have been determined with more accurate than in previous works, and the composite lightcurves have been built for the EW stars. Using relation between a period of system and an absolute magnitude obtained by Mateo and Rucinski (2017), we have calculated the absolute magnitudes of the EW stars and distances to them. The parallaxes obtained from our data differ from those given in Gaia DR 3. It may be due to insufficient quality calibration of the absolute magnitude-period relation and with the estimations of interstellar extinction. It should be noted that the photometric distances given in Gaia ESA Archive (https://gea.esac.esa.int/archive/) also differ from the astrometrical parallaxes for these stars (see section 2.2). The obtained data will be used in future for modeling kinematical and dynamical properties of observed stars.

**Acknowledgments**

The authors express their gratitude to all those people, who defend Ukraine and thus made it possible to prepare this article. This research was partly supported by the Ministry of Education and Science of Ukraine. This work has made use of data from the European Space Agency (ESA) mission Gaia (http://www.cosmos.esa.int/gaia), processed by the Gaia Data Processing and Analysis Consortium (DPAC, http://www.cosmos.esa.int/web/gaia/dpac/consortium). This research has made use of the International Variable Star Index (VSX) database, operated at AAVSO, Cambridge, Massachusetts, USA.

Table 1. Coordinates, types, magnitudes, and colors of the observed variable stars

| Gaia DR 3 | $\alpha_{2000}$ | $\delta_{2000}$ | Type | $R$ | $V-R$ |
|---|---|---|---|---|---|
| 284467709926506240 | 02 50 51.985 | +11 36 21.03 | UV Cet | 15.30 ± 0.05 | - |
| 1994611232070701184 | 23 46 51.834 | +54 59 00.17 | EW | 14.80 ± 0.02 | 0.50 ± 0.02 |
| 1997506143105623808 | 23 36 53.921 | +55 23 06.61 | EW | 15.15 ± 0.02 | 0.48 ± 0.03 |
| 1997847472748694272 | 23 37 50.051 | +56 35 34.99 | EW | 14.77 ± 0.02 | 0.56 ± 0.03 |

Table 2. Physical characteristics of EW stars.

| Gaia DR 3 | $P$ (day) | Ampl. (mag) | Gaia parallax (mas) | $M_V$ (mag) | $A_V$ (mag/kpc) | Parallax (mas) |
|---|---|---|---|---|---|---|
| 1994611232070701184 | 0.36972 ± 0.00005 | 0.39 ± 0.03 | 0.633 ± 0.068 | 4.01 ± 0.04 | 0.83 ± 0.05 | 0.862 ± 0.022 |
| 1997506143105623808 | 0.408354 ± 0.000015 | 0.42 ± 0.03 | 0.465 ± 0.027 | 3.63 ± 0.07 | 0.90 ± 0.04 | 0.714 ± 0.017 |
| 1997847472748694272 | 0.3381355 ± 0.0000015 | 0.32 ± 0.03 | 0.745 ± 0.032 | 4.34 ± 0.03 | 1.01 ± 0.03 | - |